# Effects of the Artificial Skin's Thickness on the Subsurface Pressure Profiles of Flat, Curved, and Braille Surfaces

John-John Cabibihan[1], Sushil Singh Chauhan[2], and Shruthi Suresh[2]

[1]Mechanical and Industrial Engineering Department, Qatar University

[2]Electrical and Computer Engineering Department, National University of Singapore

Corresponding author's email: john.cabibihan@qu.edu.qa

*Abstract*—**The primary interface of contact between a robotic or prosthetic hand and the external world is through the artificial skin. To make sense of that contact, tactile sensors are needed. These sensors are normally embedded in soft, synthetic materials for protecting the subsurface sensor from damage or for better hand-to-object contact. It is important to understand how the mechanical signals transmit from the artificial skin to the embedded tactile sensors. In this paper, we made use of a finite element model of an artificial fingertip with viscoelastic and hyperelastic behaviors to investigate the subsurface pressure profiles when flat, curved, and Braille surfaces were indented on the surface of the model. Furthermore, we investigated the effects of 1, 3 and 5 mm thickness of the skin on the subsurface pressure profiles. The simulation results were experimentally validated using a 25.4 µm thin pressure detecting film that was able to follow the contours of a non-planar surface, which is analogous to an artificial bone. Results show that the thickness of the artificial skin has an effect on the peak pressure, on the span of the pressure distribution, and on the overall shape of the pressure profile that was encoded on a curved subsurface structure. Furthermore, the flat, curved, and Braille surfaces can be discriminated from one another with the 1 and 3 mm artificial skin layers, but not with the 5 mm thick skin.**

## I. Introduction

For grasping, manipulation or haptic exploration to be successful, robotic or prosthetic hands would need to make sense of the world through tactile sensors. Over the years, various tactile sensing technologies and designs have been developed (see reviews in [1-5]). Embedding these tactile sensors in soft, synthetic skins have been reported to give many advantages. Among these include skin compliance for better robotic grasping and manipulation [6-9]; skin conformance for curvature, shape or object recognition [10-13]; adding fingerprint-like designs on the skin to localize contact information for roughness or texture detection [14-17]; and soft skin properties for social acceptance in human-robot interaction [18-20]. Given these benefits, it is evident that the mechanical behaviour of the artificial skins should be understood for a fundamental reason: the artificial skin is the primary interface of contact between the robotic or prosthetic finger and the external environment.

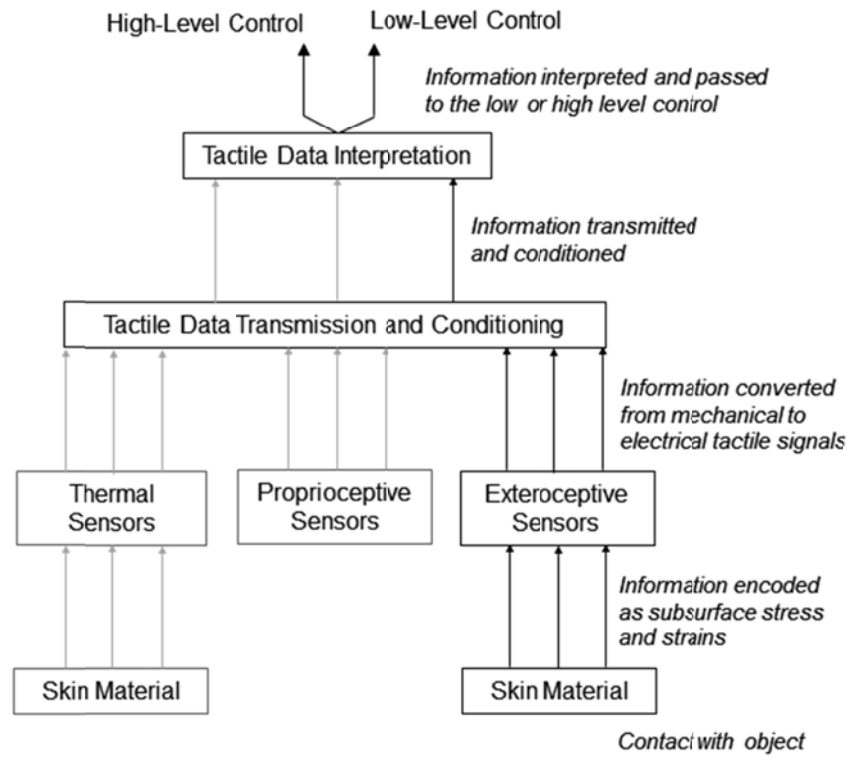

Fig. 1. General scheme of an artificial tactile sensory system. The exteroceptive branch of the system (in bold lines) may consist of four levels. The information being passed are described beside the arrows. As the tactile system gets in contact with the object or surface, information flows to the following modules: skin material, tactile sensors, data transmission and conditioning, data interpretation and finally to the low-level or high-level control modules of the robotic or prosthetic hand system.

Upon contact with any object or surface, the mechanical signals are transmitted from the skin surface to the embedded tactile sensors. Fig. 1 shows the general sequence of information flow as an artificial tactile sensory system comes in contact with an object. The scope of this paper is limited to the relationship of the artificial skin and exteroceptive sensing. Exteroceptors are used for sensing the interactions on the hand-object-environment, like force sensors that are embedded deep into the robotic finger (i.e. intrinsic sensors [21-23]) and sensor arrays that are closer to the skin surface (i.e. extrinsic sensors [24-26]). The low-level tactile encoding may typically involve two other transduction fields: proprioceptive sensors for sensing joint angles and torques (e.g. [22, 23, 27]) and thermal sensors for detecting temperatures at the surface of contact (e.g. [28, 29]).

In a bottom-up manner, the object makes contact with the skin leading to a distribution of forces on the skin surface. Second, the surface pressure or surface deflection induces subsurface stresses and strains. Third, the embedded tactile sensors detect these mechanical signals and convert them into electrical signals. Fourth, these signals are transmitted and conditioned through the appropriate wiring design, multiplexing and filtering schemes. Lastly, the main processing hardware interprets these signals and passes them to the low-level and high-level control system of the artificial hand. Significant development efforts have focused on the transducers, whereas the whole tactile system involves the equally important modules of the skin, data transmission, conditioning and data interpretation.

In this paper, we investigate the effects of the thickness of the artificial skin layer in the pressure readings at the subsurface of the artificial skin. Selecting the thickness of the skin material is important for discriminating the shape that is in contact at the skin surface through the embedded tactile sensors. The information encoded between the skin material and exteroceptive sensors are important because these are the crucial information that have to be interpreted for high level or low level control that the system will use to perceive shapes or to properly grasp or manipulate objects. The next section presents the background on previous methods. Sec. III describes the procedures used for the simulation. Sec. IV describes the experimental methods used for verifying the simulation data. Sec. V presents the results and discussion and lastly, Sec. VI gives the concluding remarks.

## II. BACKGROUND

Tactile shape recognition can be classified to be in the domain of the so-called *tactile inversion problem* [4, 17, 30, 31]. This problem can be generally described as follows: *Given the measurement of the stress or strain field at a discrete number of points on the subsurface of a skin medium, decode the contact pressure or deflection on the surface* [32]. For investigating the decoding problem, the encoding techniques reported in the literatures make use of (i) actual sensor data, (ii) simulation data or (iii) both. Finite element modeling is one of the preferred simulation techniques. A concise review based on the literature analysis in [33] is given to present a general description of the other methods for tactile data encoding:

*Elastic half-space models.* Contact analysis problems are frequently simplified by approximating the fingertip as a linear, perfectly elastic half-space [34]. Furthermore, plane strain assumptions are typically made, which means that there is no variation in the surface shape and contact conditions along the entire width of the sensor. This approach has the advantage of analytical tractability and offers fairly good approximations despite the simplifications made [33]. For example, the elastic half-space model was used to obtain the location and amplitude of a normal load experienced by a singled tactile element from the measurements of the three local strain components at one point under a flat covering [17].

*Transfer functions* provide another solution. With this approach the fundamental differential equations of equilibrium and the stress-strain relationship equations (generalized Hooke's law) are transformed using, for instance, the two-dimensional Fast Fourier Transform, the Walsh-Hadamard Transform, or the Direct Cosine transform, over the (x, y) space [35]. These transforms have properties of shift invariance whereby an object placed anywhere on the sensor array will yield the same transform that is independent of the object's orientation on the sensor array. Accordingly, template matching can be performed [35] without shifting the reference object through numerous possible positions and orientations that the object can have on the sensor array.

The *FE method* makes it possible for arbitrary objects, sensor geometries, and loading conditions to be modeled, which is difficult for the elastic half-space approach. Real fingertips have geometries that are not exactly half-spaces, are typically inhomogeneous due to the embedded sensors, and have material properties with viscoelastic components making the behavior obviously nonlinear [30]. This paper used the FE method as a step towards understanding the role of the artificial skin's thickness to discriminate various shapes from the subsurface information.

Using a linear elastic FE model of a planar skin, it has been observed in [36] that a skin covering effectively blurs the signals from a probe with an indenting tip of 3 mm radius transmitted to the embedded sensors at 1 mm depth. Furthermore, FE models were used to examine the stresses and strains beneath finger ridge-like structures with an elastic hemisphere model in [17, 37] and a hyperelastic-viscoelastic model in [38]. The current paper investigates the subsurface effects of different surfaces that were indented on the skin surface of silicone and polyurethane materials of various skin thicknesses.

Unlike several of the previous studies [12, 36, 39, 40], the simulation conditions presented in the current paper better represent realistic conditions; soft skin materials for tactile sensing are generally viscoelastic [41]. The FE model used here is viscoelastic-hyperelastic and was implemented on an external cross-sectional geometry similar to a human fingertip. By evaluating the variations in the resulting subsurface pressure profiles, the optimal skin thickness can be selected.

## III. SIMULATION METHODS

### A. Simulation Procedure

An artificial finger model that consists of the fingertip skin, bone, a stiff tissue, and nail is shown on Fig. 2A. For computational efficiency, only half of the fingertip skin's cross section was modeled (i.e. the meshed geometry).

A two-dimensional finite element model of a fingertip was created in the commercial finite element software Abaqus™/Standard 6.8-EF (Dassault Systemes Simulia Corp., Providence, RI, USA). The plain strain 8-node bi-quadratic element type was used to model the fingertip. A thickness unit of 1 was set (measurement towards the page). The interface at the bottom region of the model (i.e. the bone, stiff tissue, and nail region) was fixed in all the degrees of freedom. The geometry approximates the stiff bone, nail, and the tissue between them. From the x-ray analysis in [42], this tissue can be assumed to be stiff. Three fingertip models with 1, 3 and 5 mm thickness were then simulated.

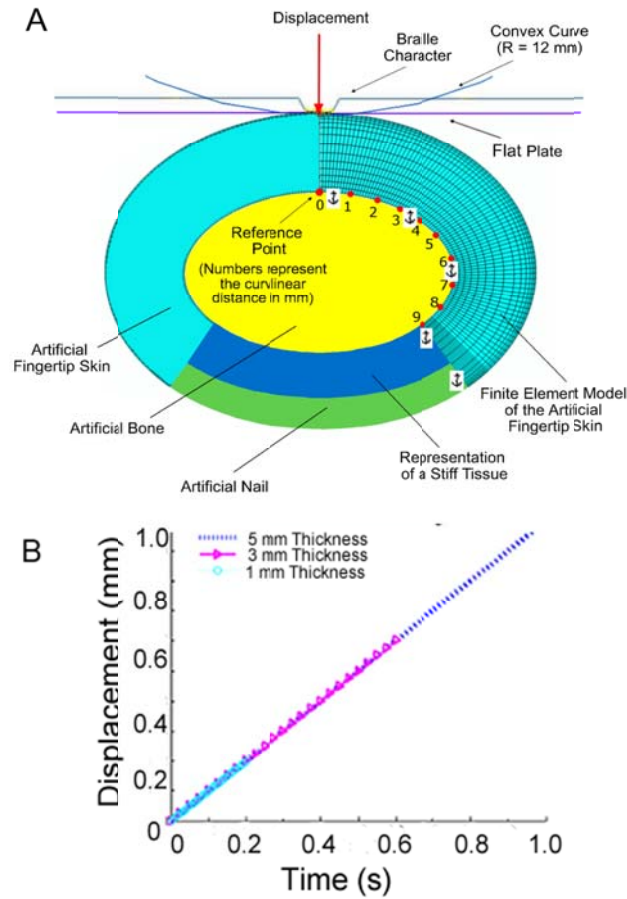

Fig. 2. Artificial fingertip model and simulation conditions. (A) The model of an artificial fingertip consisting of an artificial skin, bone, and nail. Only the meshed geometry was simulated for computational efficiency. The fingertip model also shows the locations where the plate, convex curve, and Braille character were applied. The numbers at the surface of the artificial bone represent the curvilinear distance in mm. The artificial skin was fixed at all degrees-of-freedom at the skin's interface with the bone, stiff tissue, and nail surface. (B) The applied displacement-controlled indentation, which is equivalent to 20% nominal strains of the 1, 3 and 5 mm skin thicknesses.

The indenters were set to have a vertical downward movement towards the fingertip. A flat surface, a convex curve with a 12 mm radius, and a cross-section of a Braille character were indented. The geometry of the Braille character was based on the technical specifications of the Standard American measurements (www.tiresias.org/ reports/braille_cell.htm), which follows the convention wherein a height of 0.48 mm and a base diameter of 1.44 mm is maintained. The three surface types were indented with 20% nominal strain of the fingertip model's thickness corresponding to 0.2, 0.6 and 1 mm indentations for the 1, 3 and 5 mm skin thickness, respectively. The rate of indentation was 1 mm/s (Fig. 2B). Data for each of the skin thicknesses were obtained at the final time step.

It was suggested that the minimal independent sensors for tactile signal decoding should be the mean pressure and two components of shear [33]. For normal applied indentation of the surface shapes, the mean pressure should be sufficient. Hence, the nodes at the artificial bone surface were selected, with the starting node for the path shown on Fig. 2A as the 0 reference point. The mean pressure output for these nodes along the bone surface was plotted. The mean pressure (also known as mean normal stress) is equivalent to

$$P = \frac{1}{3}(\sigma_x + \sigma_y + \sigma_z) \qquad (1)$$

where $\sigma$ denotes the normal stresses on those axes and where there are no shear stresses involved [43].

*B. Material Samples Used*

The materials that have been used in earlier works as soft skin materials for prosthetic and robotic fingers [18, 44-46] were evaluated in this study. These were: silicone (GLS 40, Prochima, s.n.c., Italy) and polyurethane (Poly 74-45, Polytek Devt Corp, USA). The silicone sample has a Shore A value of 11 while the polyurethane sample has a value of 45, where a lower value indicates a low resistance to an indenter in a standard durometer test.

*C. Material Models*

The synthetic skins were assumed to behave with hyperelastic and viscoelastic properties [18, 19]. As such, the total stress was made equivalent to the sum of the hyperelastic (HE) stress and the viscoelastic (VE) stress such that:

$$\sigma(t) = \sigma_{HE}(t) + \sigma_{VE}(t) \qquad (2)$$

where $t$ is the time. A strain energy function, $U$, defined in Storakers [47] for highly compressible elastomers was used to describe the hyperelastic behaviour of the synthetic materials. The model was likewise implemented in simulation studies of a human fingertip in [48]. The function is given as:

$$U = \sum_{i=1}^{N} \frac{2\mu_i}{\alpha_i^2} \left[ \lambda_1^{\alpha_i} + \lambda_2^{\alpha_i} + \lambda_3^{\alpha_i} - 3 + \frac{1}{\beta}(J^{-\alpha_i\beta} - 1) \right] \qquad (3)$$

where $\mu_i$ denote the shear moduli, $\alpha_i$ are dimensionless material parameters, $\lambda_i$ are the principal stretch ratios, $J = \lambda_1 \lambda_2 \lambda_3$ is the volume ratio and $N$ is the number of terms used in the strain energy function. The coefficient $\beta$ determines the degrees of compressibility in the energy function. The relationship of $\beta$ to the Poisson's ratio, $\upsilon$, is $\beta = \upsilon/(1 - 2\upsilon)$.

The hyperelastic stress is related to the strain energy function (Eqn. 3) by:

$$\sigma_{HE} = \frac{2}{J} F \frac{\partial U}{\partial C} F^T \qquad (4)$$

where $F$ is the deformation gradient and $C$ is the right Cauchy-Green deformation tensors.

The viscoelastic behavior is defined below, with a relaxation function *g(t)* applied to the hyperelastic stress:

$$\sigma_{VE} = \int_0^t \dot{g}(\tau)\sigma_{HE}(t - \tau)d\tau \qquad (5)$$

The viscoelastic material is defined by a Prony series expansion of the relaxation function [49]:

$$g(t) = \left[ 1 - \sum_{i=1}^{N_G} g_i(1 - e^{-t/\tau_i}) \right] \qquad (6)$$

where, $g_i$ is the shear relaxation modulus ratio, $\tau_i$ is the relaxation time, and $N_G$ denotes the number of terms used in the relaxation function. The detailed information on how the governing equations are numerically solved are described in the Abaqus/CAE Theory Manual [50].

Table I at the Appendix shows the coefficients for silicone and polyurethane. These material parameters were earlier validated in [18]. The validation procedures in that work consisted of having the simulation results in the finite element models and compared with the experimental results from the physical samples of synthetic fingers that were made of silicone and polyurethane materials using imaging and indentation tests. The results from simulation and validation experiments were in good agreement.

## IV. Experimental Procedures

### A. Materials and Equipment

To validate the simulation results, the materials and equipment used in the experiment were as follows:

1) *Surface Pressure Mapping Sensor*: The Pressurex-micro (Sensor Products, Inc., USA) is a tactile pressure indicating sensor film that can determine the relative pressure between two surfaces in contact. Pressurex-micro is a thin pressure sensor (20 mils or 25.4 μm) and can be used in tight spaces where conventional electronic transducers cannot be used. This film activates with contact pressures that are less than 1.5 kg/cm$^2$. The pressure film irreversibly records the increase in pressure. The resulting image shows the relative amount of pressure as a grayscale distribution profile where the higher the pressure applied results into a darker gray level.

2) *Universal Testing Machine:* A universal testing machine (MicroTester Model 5848, Instron) was used to calibrate the Pressurex-micro film and to apply the loads on the fabricated skins. The data were acquired using the built-in software (Merlin software). Attached to the movable column is the indenter (Fig. 3A).

3) *Indenters:* The indenters in Fig. 3B were fabricated in accordance to the simulation geometries in Fig. 2A. The curved and Braille surfaces were 3D printed (Model 350, Objet Geometries Ltd., USA) with acrylic-based photo-polymer materials (Fullcure VeroWhite, Code 830, Objet Geometries Ltd., USA).

### B. Calibration of the Pressure Detecting Film

In order to ensure the accurate analysis of the results from the pressure film, a separate calibration procedure was performed to obtain the relationship between the pressure applied and the mean grayscale value of the pressure film. The mean grayscale value is a representation of the average value of the grayscale value of the pixels found on the image.

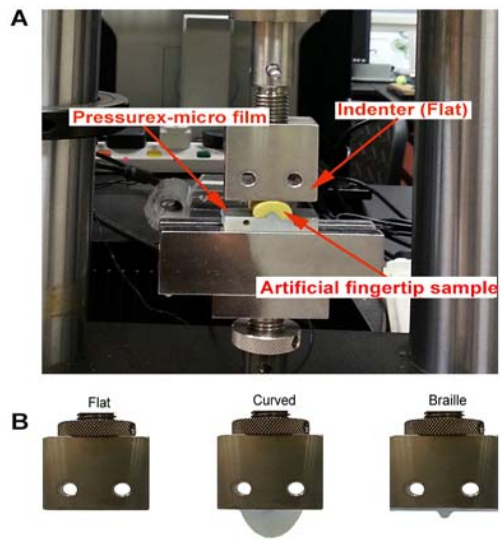

Fig. 3. Experimental set-up. (A) Indention of an artificial fingertip sample. The Pressurex-micro film was mounted along the stiff base and between the artificial fingertip skin and the representation of an artificial bone structure. (B) The flat, curved, and Braille surfaces used in the indentation experiments.

The calibration tests were conducted by compressing a controlled load ranging from 1 to 30 N at the rate of 1 mm/s. The calibration tests were performed using the testing machine by the application of the loads on the pressure detecting film. The pressure at the contacting surfaces between the aluminum cylinder and the sensor film for the experiments of calibration were calculated based on Eqn. (7).

$$P = \frac{F}{A} \qquad (7)$$

where $P$ is the pressure in N/mm$^2$ (MPa), $F$ is the force applied in N and $A$ is the area of the surface in mm$^2$. The area of the aluminum cylinder's surface was 100 mm$^2$.

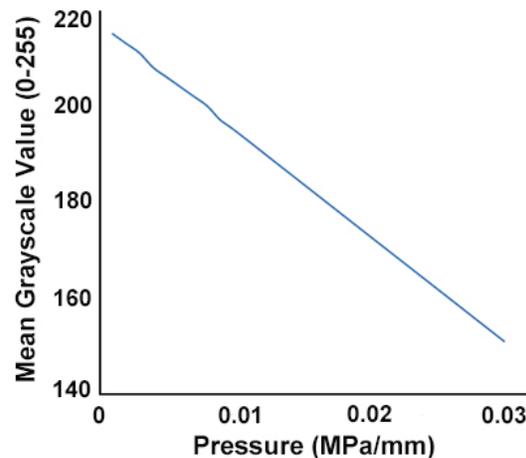

Fig. 4. Relationship of the mean grayscale value and applied pressure of the Pressurex-micro film.

The resulting trend line is shown in Fig. 4 to represent the relationship between the pressure and the mean grayscale value. By evaluating the statistical errors resulting from a different type of fitting function, a linear fitting type can best describe the relationship. The function, with the 90% confidence bounds, is given as:

$$P_i = -0.00425 \times GS + 0.978 \qquad (8)$$

where, $P_i$ is the pressure per unit length in MPa/mm and $GS$ denotes the mean grayscale value.

For the pressure detecting film that was embedded in the artificial skins, the pressure per unit length (in MPa/mm) was achieved by dividing Eqn (7) by the length Y (Fig. 5). Y is a representation of the length along which the indentation was done (10 mm), while X is the length (20 mm) which was kept constant throughout the experimentation. When the pressure detecting film was scanned, the pixel ratio X:Y obtained was 160:80 pixels. As these values of X and Y correspond to 20 mm and 10 mm, we can deduce that 1 mm corresponds to 8 pixels of the scanned image. It is to be noted that the value of $x_i$ varies with the different indenters used and the thickness of the skin. The values of mean grayscale as well as pressure per unit length as seen in Fig. 5 were based on this ratio.

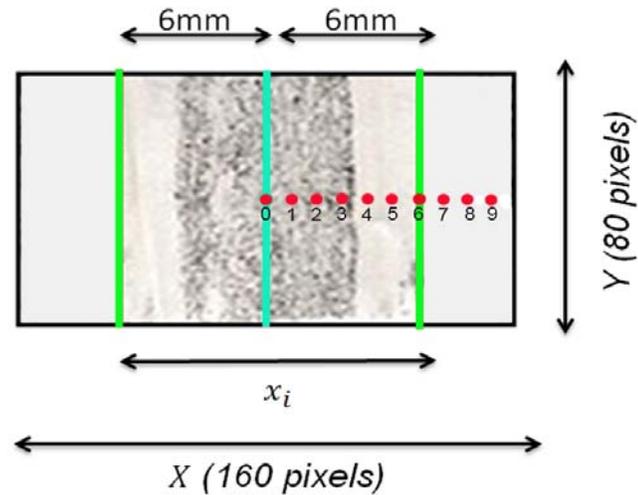

Fig. 5. The resulting image on the pressure detecting film after the indentation experiment on a 5 mm silicone sample using a curved indenter.

## C. Indentation of the Embedded Pressure Films

The indenters were mounted on the testing machine and were aligned with the surface of the artificial skin. In accordance to the simulation conditions, all fingertip samples were then indented with 20% nominal strain of the thickness, which corresponds to 0.2, 0.6 and 1 mm indentations for the 1, 3 and 5 mm skin thicknesses, respectively. The indentation velocity was set to 1 mm/s. The indentation results in grayscale are shown in Fig. 6.

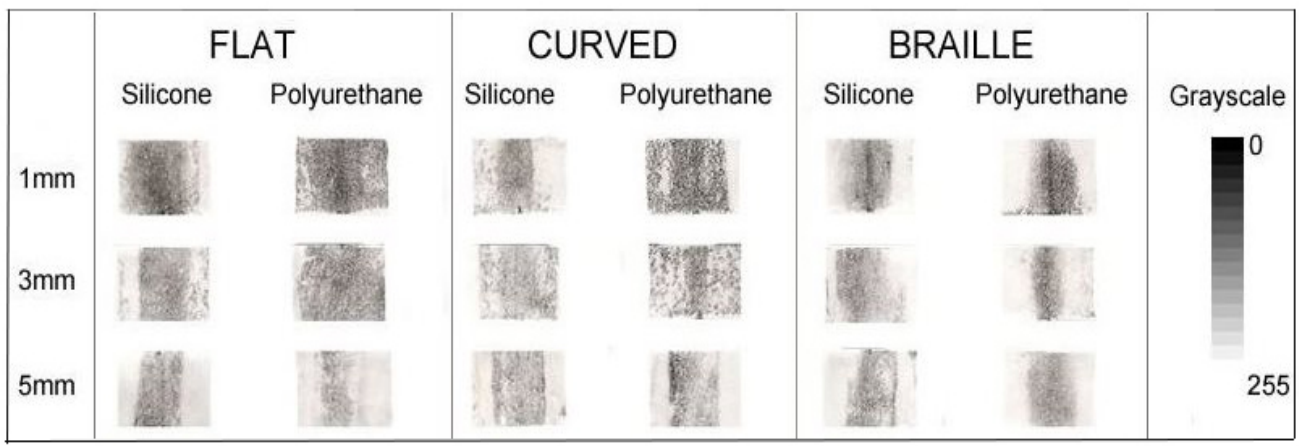

Fig. 6. Indentation experiment results showing the grayscale images generated from the full strip of the pressure detecting films.

## D. Software Development

In order to analyze the subsurface pressure distribution, scans of the pressure films were acquired using a scanner (HP Scanjet 200 Scanner, Hewlett-Packard, USA) and were analyzed with a software that was developed on a computing package (MATLAB, R13a, MathWorks, Inc., USA). Fig. 7 shows a screenshot of the customized software. It has the following features: (i) it allows the analysis of individual pixels; (ii) it obtains the grayscale value of the individual pixels; and (iii) it converts the grayscale value into pressure values using Eqn. (8).

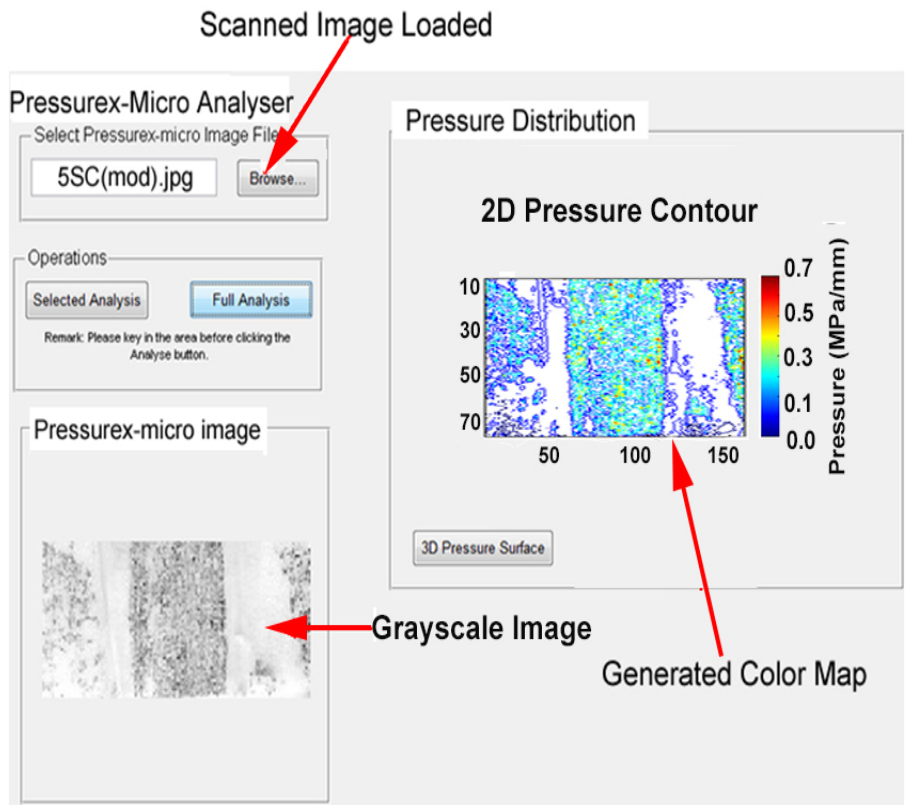

Fig. 7. Screenshot of the graphical user interface to convert the grayscale images to color maps.

The imaging software represents each pixel in terms of a grayscale value, which ranges from 0 to 255, where 0 is the representation of a black pixel and 255 is the representation of a white pixel. These pressure values were then expressed as color maps. The color map is an m-by-3 matrix of real numbers between 0.0 and 1.0 and each row is an RGB vector, which defines a color. A built-in MATLAB function, *jet*, was used to create a vector of *n* colors beginning with dark blue, ranging through shades of blue, cyan, green, yellow and red, and ending with dark red.

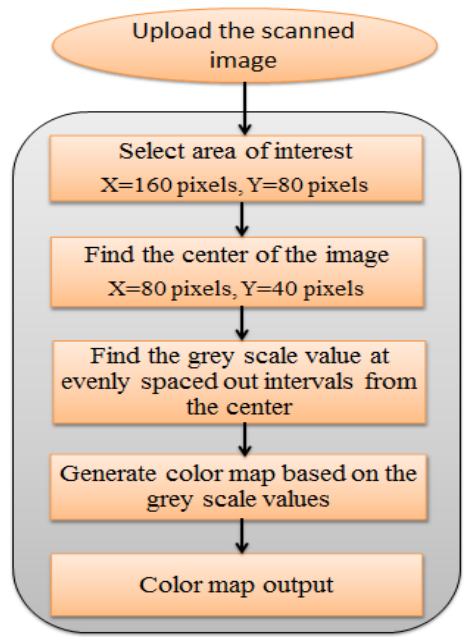

Fig. 8. Algorithm for the analysis of pressure film to generate color maps.

The scanned images of the pressure films were uploaded to the software in order to provide an analysis of a particular section of the pressure detecting film. Once the selection has been made, the software runs the algorithm in Fig. 8 and provides an output of color maps and pressure values. The analysis also presents a color bar that indicates the pressure denoted by each individual color range.

## V. RESULTS AND DISCUSSION

### A. Pressure Contours from Simulations and Experiments

The pressure contours from the FE simulations are shown in Fig. 9A. The color variations were concentrated on the contact surface. The mechanical signals were transmitted downwards toward the fixed region at the bottom of the fingertip model, which can be thought of as a bone structure. The curvilinear lengths at the subsurface of the 1, 3 and 5 mm thick skins were measured to be 14, 9 and 5 mm, respectively. From the simulation results, the pressure at the subsurface occurs within 6 mm of the curvilinear length. Fig. 9B shows the experimental results when the pressure detecting films were mounted between the artificial skin layers and the curved surface of a rigid structure. Because the effects of pressure were known from simulations to occur only within the 6 mm of curvilinear length, we only show the results of the pressure detecting film to be at that length (please also refer to Fig. 5 for the illustration).

The pressure data from simulation and experiments were then plotted in Fig. 10. The next subsections describe the effects of the thickness on the peak pressure, on the span of the pressure distribution, and on the overall shape of the pressure profile.

### B. Effect on the Peak Pressures

For both the simulation and experimental results, the large magnitudes of pressure can be observed at the central regions of the artificial fingertip (Fig. 10). When plotted, we can see in Fig. 11 that in general, the flat surface registered the largest peak pressures, followed by the curved and Braille surfaces for the 1 mm and 3 mm thick skins, but not with the 5 mm thick skin.

As the skin thickness increases, the amount of pressure reduces in magnitude. In other words, the skin thickness dampens the amount of pressure that was transmitted downwards to where the sensors could be possibly mounted. On one hand, a thicker skin can protect the subsurface sensors because the skin absorbs the pressure that reaches the sensor. On the other hand, the pressure signals can be very small to be appropriately detected.

To validate the simulation results, we plotted the simulation data beside the experimental data. Fig. 11A shows the results for silicone, while Fig. 11B shows the results for polyurethane. Notice that the pressure magnitude per unit length is lower for silicone than in polyurethane. The reason is that silicone is softer than polyurethane and it takes less effort to indent the silicone material according to the desired displacement. For example, for a 1 mm thick skin with a flat indenter, the peak pressure for silicone was about 0.5 MPa/mm (Fig. 11A) while it was about 0.8 MPa/mm for polyurethane (Fig. 11B).

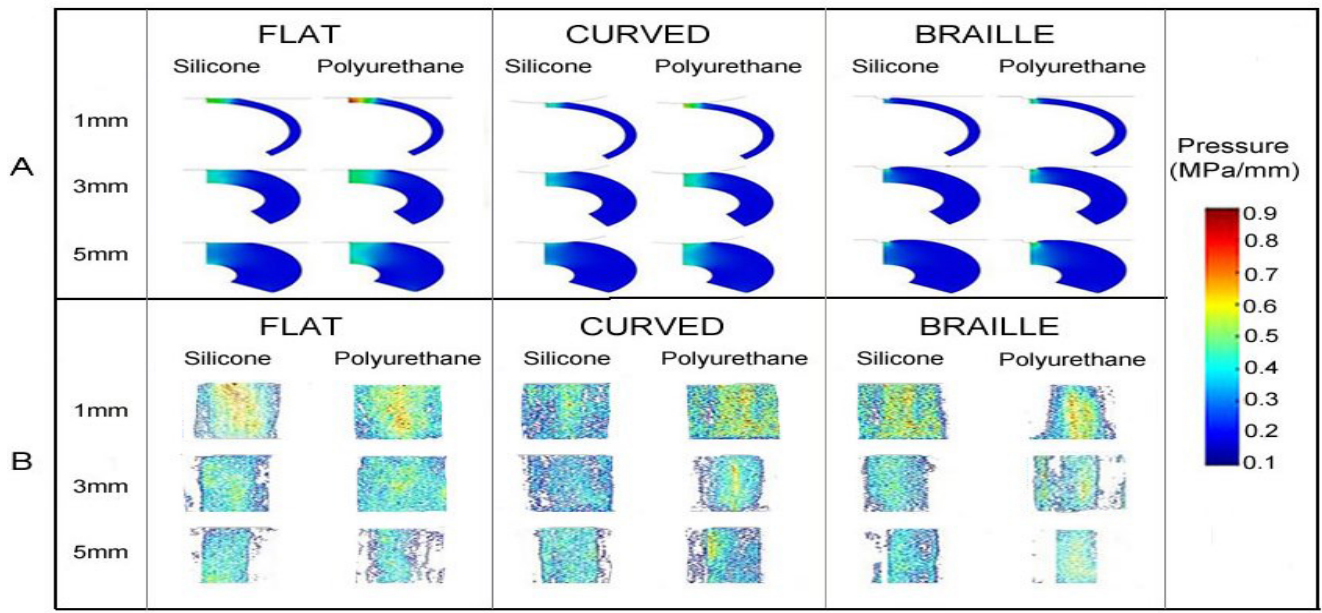

Fig. 9. Simulation and indentation experiment results. (A) simulation contours from the finite element analysis showing half of the fingertip model, and (B) color maps generated from grayscale images showing the strip of the pressure detecting films.

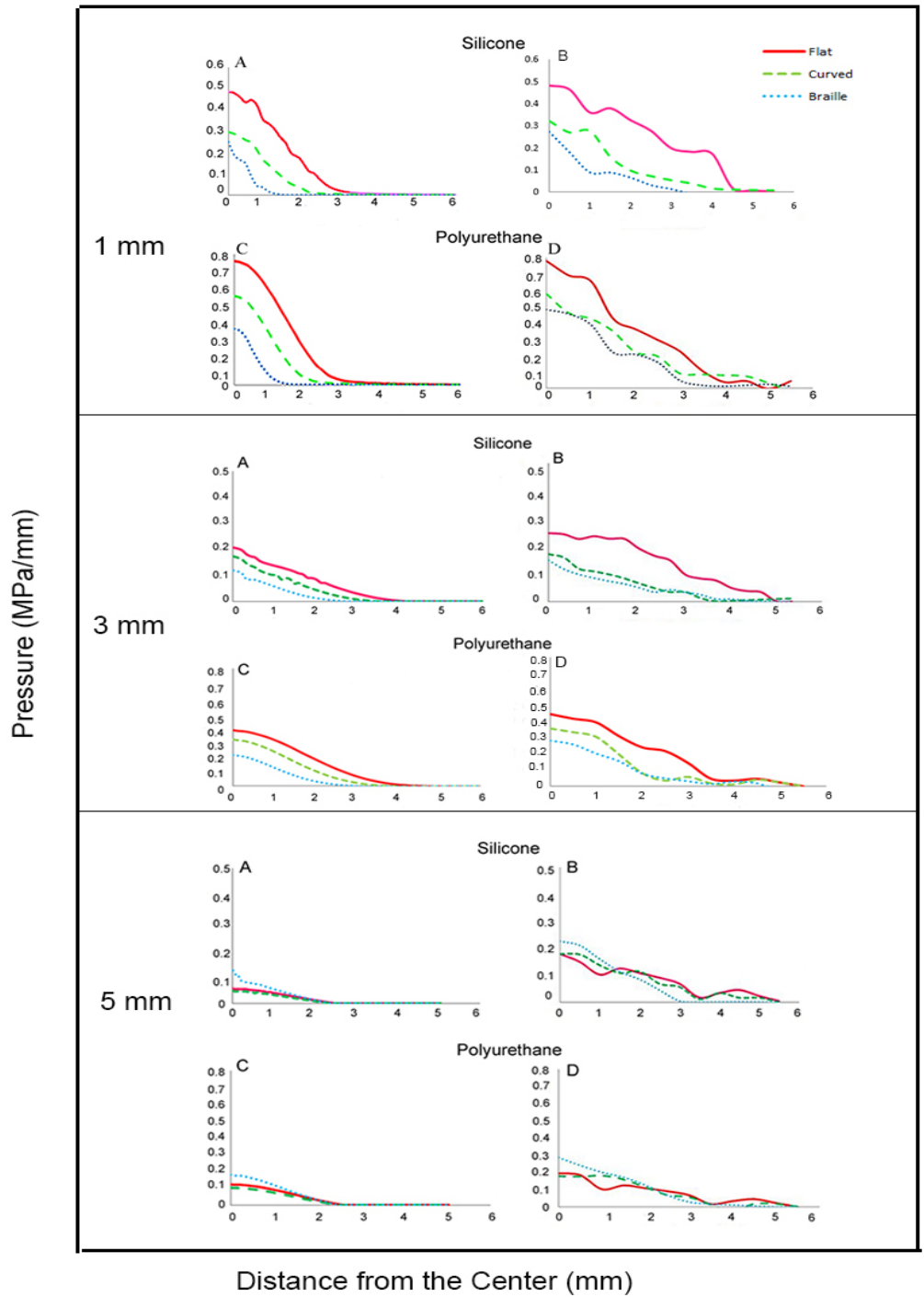

Fig. 10. Pressure contours of 1 mm, 2 mm and 5 mm skins compared with data from simulations for the silicone and polyurethane samples. A and C are simulated results while B and D are the experimental results.

For the 1 mm thick skin for silicone, the percentage differences for simulation and experiments were found to be 1.3%, 8% and 9% for flat, curved, and Braille indenters, respectively. The percentage differences for polyurethane for simulation and experimentation were found to be 2%, 1.4% and 33% for flat, curved, and Braille indenters, respectively. For the 3 mm thickness, the percent differences of the peak pressures for silicone in the simulation and experiments were 19%, 0% and 40% for the flat, curved and Braille indenters, respectively. For polyurethane, the differences were 18%, 13%, and 5%. For the 5 mm thickness, the simulation versus experimental differences in the results for silicone material were 231%, 292% and 87%, for flat, curved and Braille, respectively while for polyurethane, the differences were 59%, 73% and 59% for the same sequence of the indenter shapes.

The simulation model made use of plane strain elements. In using this type of element, the assumption is that the constant deformations and strains occur at the axis towards the page ([43]; see Fig. 2A). On the other hand, experimental results on the incompressible materials, like the silicone and polyurethane that were used in the current paper, will have deformations and strains in the directions that were not constrained. The differences in simulation and experiments became pronounced in the peak pressures of the 5 mm thickness in silicone whereby more than 200% differences were observed. Recall that each skin thickness type was indented with 20% nominal strain of the fingertip thickness (i.e. 1, 0.6 and 0.2 mm indentations for the 5, 3 and 1 mm skin thickness, respectively). Although large differences were observed, the general trend for both simulation and experiment is according to decreasing peak pressures for Braille, flat, and curved indenters.

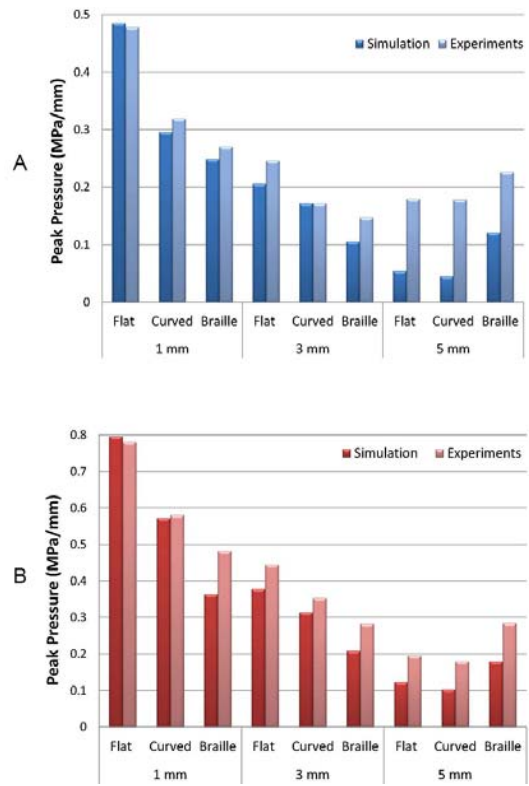

Fig. 11. Peak pressure variation for the 1, 3 and 5 mm (A) silicone and (B) polyurethane skin samples. These were indented with flat, curved, and Braille surfaces.

## C. *Effect on the Span of the Pressure Profile*

The thickness of the artificial skin has an effect on the span of the pressure distribution along the subsurface. As the skin becomes thicker, the span of the pressure distribution becomes smaller. In general, for the 1 mm and 3 mm thick skins for both the silicone and polyurethane materials, the flat indenter has the largest span, followed by the curved and lastly by the Braille (Fig. 12). For the 5 mm thick skins, there were minimal differences on the span among the three indenter types. Because the skin was too thick, the representations of flat, curved and Braille were already indistinguishable at the subsurface. For both silicone and polyurethane, for example, the distances from the center of the 5 mm thick skins were within 2.3 to 2.5 mm for the simulations and within 3 to 3.5 mm for the experiments for the various indenter types.

The information on the span of pressure distribution would be very helpful in determining the minimum elements of a sensory array when they are embedded on an artificial skin. Fearing and Hollerbach [12] suggest that the Nyquist sampling theorem can be applied to determine the density of the sensor elements to detect a pattern of tactile signals. The theorem states that any bandlimited signal can be recovered from the discrete samples fully if the samples are taken at sufficiently high frequency. This sampling frequency should be at least twice the maximum signal frequency contained in the signal and this can be taken as the minimum element-to-element distance of the embedded sensors of negligible width to recover the continuous measurements from discrete samples. To illustrate, the reader is referred to the simulation results of the 1 mm skin thickness for silicone in Fig. 10. To be able to recover the signals of a Braille indenter from pressure readings of a subsurface sensor, a 4 mm element-to-element sensor distance will be too large to achieve a pressure signal that represents a Braille surface.

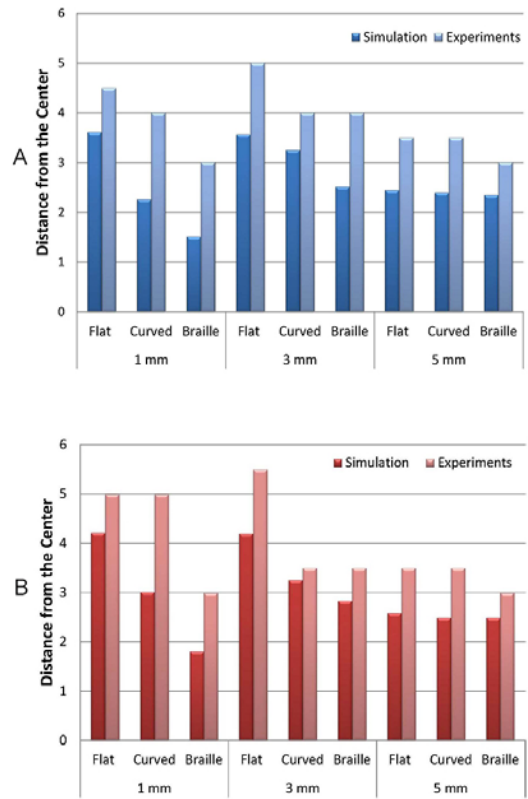

Fig. 12. Span of the pressure distribution for the 1, 3 and 5 mm (A) silicone. and (B) polyurethane skin samples. These were indented with flat, curved, and Braille surfaces.

The span from the Braille character is 1.5 mm. Hence, a minimum element-to-element distance of 0.75 mm would be able to give a satisfactory spatial resolution. In humans, the spatial resolution was measured to be 0.87 mm using the gap detection test [51].

*D. Effect on the Shape of the Pressure Profile*

The thickness of the artificial skin has an effect on the overall shape of the pressure profiles. Fig. 10 shows that the resulting pressure profiles from the plate, curved, and the Braille character can be discriminated from one another for both the silicone and polyurethane materials for the 1 mm and 3 mm thick skins. For the 5 mm-thick skins, the profiles of the plate and the curved surface have already overlapped. The full height of 0.48 mm of the Braille character has completely indented the skin. Regardless, the edges of the Braille character could not be seen in the silicone or polyurethane results for all the three thickness types.

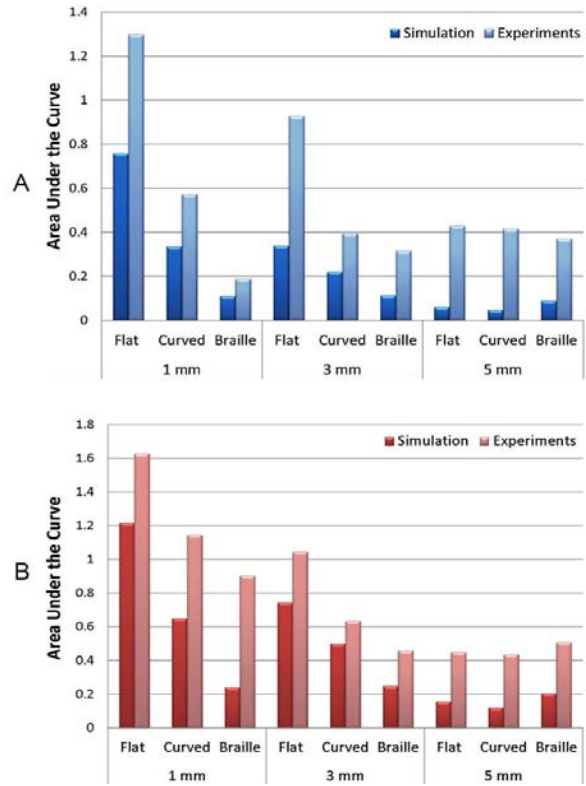

Fig. 13. Area under the curve from the simulation data of the 1, 3 and 5 mm for (A) silicone and (B) polyurethane skin samples. These were indented with flat, curved, and Braille surfaces.

A skin layer of less than 1 mm would be able to show the important features of a Braille character for tactile recognition. Furthermore, we compared the area under the curve (AUC) for each of the resulting pressure profiles. Such information can be useful for discriminating one surface from another. From Fig. 13, distinct differences in the AUC can be observed in the 1 mm and 3 mm thicknesses for silicone and polyurethane skins. For instance, referring to the experimental results of silicone with 1 mm thickness, the AUC was found to be 1.3, 0.6 and 0.2 for the flat, curved and Braille, respectively. However, for the 5 mm thick skins, the AUC was nearly similar for flat, curved, and Braille at about 0.4.

Discriminating among the shape of the indenters can take advantage of machine learning techniques. For example, Naïve Bayes (NB), Artificial Neural Networks (ANN), and Support Vector Machines (SVM) techniques were used to discriminate among flat, curved and edged indenters using a 2×2 element tactile sensor array [11].

## VI. Conclusion

The fingertip-object interface is the best location to give the information regarding critical properties and events at the region of contact. Through touch, it is possible to directly measure the physical properties of an object such as shape, hardness, temperature, vibration, mass and surface details. Alternatively, computer vision, through cameras, can detect such properties but they can be deduced indirectly, and this only possible when the target object is not occluded.

Analogous to how video cameras are being used to deduce object properties in computer vision, the present account provided a validated Finite Element model of an artificial fingertip that serves as a tool to visualize the internal effects from externally applied shapes. More specifically, these were the primitive surfaces of objects, such as flat, curved or corners as represented by a Braille character. With this kind of approach, we were able to evaluate the effects of different thicknesses of artificial skins on the pressure profiles underneath the surface. An alternative approach is to carry out a trial-and-error approach of embedding a tactile sensor array within a soft, synthetic skin and evaluating the subsurface mechanical signals like pressure, stress or strain.

The artificial skin's thickness has the following effects. First, the thickness dampens the amount of pressure that is transmitted to the embedded sensors, with consequences on the sensing ability and the robustness of the sensors. Second, a thicker skin reduces the span of the pressure distribution along the subsurface where the sensors will be embedded. The information on the span can give the spatial resolution of tactile sensing arrays to encode the signals from the surface. Lastly, a thicker artificial skin blurs the overall shape of the pressure signals at the subsurface, which could reduce the possibility for the discrimination of the tactile signals in machine learning implementation.

The results suggest that near-surface tactile sensors could allow better discrimination of different shapes. The skin material acts like a low pass filter [36]. Embedding the sensor with a thick skin will blur the mechanical signals being transmitted to the sensor. On the other hand, embedding the sensor with a thin skin improves its detection ability but at the same time making it vulnerable to damage. A comprehensive literature review on the tactile sensing on high-density arrays in [2] shows that microelectro-mechanical systems (MEMS) and silicon-micromachined devices are popular techniques for miniaturization. The compromise between robustness and sensitivity is an important consideration for design trade-off decisions, especially for the inherently fragile MEMS and silicon-based tactile sensors.

In consideration of the human tactile system, the mechanoreceptors known to be sensitive to surface details, shapes, and orientations are the Slowly Adapting type 1 (SA-I) and Fast Adapting type 1 (FA-I) mechanoreceptors [52-54]. They are located at about 0.7 to 1 mm below the skin surface [51]. However, for the engineered counterpart, placing the tactile sensors near the surface will make them more vulnerable to damage. Soft synthetic skins of 1 mm thickness can also get punctured easily. While a 5 mm thick skin can protect the sensors better, the results show that the mechanical signals are already blurred. A skin thickness of about 3 mm would be optimal for the tactile sensing and less vulnerable to damage. Future work can incorporate these findings for the design of tactile sensing systems of robotic or prosthetic hands for active exploration.

# APPENDIX

**Table I** – Coefficients of the Artificial Skin Materials

| $i$ | 1 | 2 | 3 |
|---|---|---|---|
| **Silicone** ($v = 0.49$) | | | |
| $g_i$ | 0.015 | 0.044 | 0.029 |
| $\tau_i$ (sec) | 0.025 | 0.150 | 0.300 |
| $\mu_i$ (MPa) | 0.080 | 0.010 | - |
| $\alpha_i$ | 0.001 | 15.500 | - |
| **Polyurethane** ($v = 0.47$) | | | |
| $g_i$ | 0.167 | 0.158 | 0.113 |
| $\tau_i$ (sec) | 0.100 | 1.380 | 25.472 |
| $\mu_i$ (MPa) | 0.100 | 0.063 | - |
| $\alpha_i$ | 5.500 | 8.250 | - |

## ACKNOWLEDGMENT

This work was supported in part by the National University of Singapore Academic Research Funding Grant No. R-263-000-A21-112. This paper is a significantly expanded version of a paper presented at the IEEE International Conference on Biomedical Robotics and Biomechatronics, Rome, Italy, June 2012. The authors thank Mr. Ng Khoon Siong of the National University of Singapore for the software development and for the preliminary experiments on the pressure detecting films.

**John-John Cabibihan** was conferred with a PhD in biomedical robotics by the Scuola Superiore Sant'Anna, Pisa, Italy in 2007. During his PhD studies, he was awarded an International Scholarship grant by the Ecole Normale Supérieure de Cachan, France, which he spent at the Laboratoire de Mécanique et Technologie in 2004. He is currently affiliated with the Mechanical and Industrial Engineering Department of Qatar University. From 2008 to 2013, he was with the Electrical and Computer Engineering Department of the National University of Singapore, where he served as the Deputy Director of the Social Robotics Laboratory and an Affiliate Faculty Member at the Singapore Institute of Neurotechnologies (SiNAPSE). Currently, he serves at the Editorial Board of the International Journal of Social Robotics, Computational Cognitive Science, Frontiers in Bionics and the International Journal of Advanced Robotics Systems. He was a past Chair of the IEEE Systems, Man and Cybernetics Society, Singapore Chapter (terms: 2011 and 2012). He has been active in conference organizations: he was the Program Co-Chair of the 2010 International Conference on Social Robotics, Singapore; Program Chair of the 2012 International Conference on Social Robotics at Chengdu, China; and General Chair of the 2013 IEEE International Conference on Cybernetics and Intelligent Systems, Manila, Philippines. He is working on the core technologies towards lifelike touch and gestures for prosthetics and social robotics.

**Sushil Singh Chauhan** received his Master's in Science degree in Electrical Engineering with specialization in Automation and Control from the National University of Singapore in 2013 and Bachelor of Technology degree in Electronics and Instrumentation from Vellore Institute of Technology, India in 2011. He worked as Research Engineer at NUS Department of Electrical and Computer Engineering for Social Robotics Lab. Currently he is working as Research Assistant at the Singapore Institute for Neurotechnology (SINAPSE) at NUS Singapore. His research interests are in the field of neural prosthesis with the implantable electronics for peripheral nerve injury (PNI) and wireless power transmission. Further, his interest also lies in physiological measurement, social robotics for autism assessment, therapy and early intervention.

**Shruthi Suresh** graduated from the National University of Singapore's Department of Electrical and Computer Engineering in 2013 with a Bachelor's of Engineering degree. She is currently pursuing research in the field of medical devices and is passionate about developing novel uses of technology in medicine.